\documentclass[12pt,a4paper]{article}    

\begin{document}

\def\ref#1{\noindent \parshape=2 0pt 442pt 8pt 434pt #1}
\def\rif#1{\noindent \parshape=2 15mm 125mm 15mm 125mm #1}

\Large

\centerline{\bf FIELD--GUIDED PROTON ACCELERATION AT}

\vskip 0.2cm

\centerline{\bf RECONNECTING X--POINTS IN FLARES} 

\vskip 0.5cm

\large

\centerline{B. Hamilton$^1$, K. G. McClements$^2$, L. Fletcher$^1$, A. 
Thyagaraja$^2$}

\normalsize

\vskip 0.5cm

\centerline{$^1$Department of Physics and Astronomy, University of Glasgow, 
Glasgow,} 
\centerline{G12 8QQ, UK}

\vskip 0.2cm

\centerline{$^2$UKAEA Culham Division, Culham Science Centre, Abingdon, 
Oxfordshire,} 
\centerline{OX14 3BD, UK}

\vskip 0.5cm

\begin{abstract}
An explicitly energy--conserving full orbit code CUEBIT, developed
originally to describe energetic particle effects in laboratory fusion 
experiments, has been applied to the problem of proton acceleration in solar 
flares. The model fields are obtained from solutions of the linearised MHD 
equations for reconnecting modes at an X--type neutral point, with the 
additional ingredient of a longitudinal magnetic field component. To 
accelerate protons to the highest observed energies on flare timescales, it is
necessary to invoke anomalous resistivity in the MHD solution. It is shown 
that the addition of a longitudinal field component greatly increases  
the efficiency of ion acceleration, essentially because it greatly reduces 
the magnitude of drift motions away from the vicinity of the 
X-point, where the accelerating component of the electric field is largest. 
Using plasma parameters consistent with flare observations, we obtain proton 
distributions extending up to $\gamma$-ray-emitting energies ($>1\,$MeV). In 
some cases the energy distributions exhibit a bump--on--tail in the MeV range.
In general, the shape of the distribution is sensitive to the model 
parameters.  
\end{abstract}

\vskip 0.2cm

\centerline{Accepted for publication in {\it Solar Physics} February 20 2003}

\vskip 0.2cm

\section{Introduction}

The process of magnetic reconnection is believed to be intrinsic to
solar flares. It is invoked as the mechanism whereby energy-loaded
magnetic fields can reconfigure to a lower energy state, liberating
the energy which powers particle acceleration, heating, and mass
motions. As the resistivity of solar coronal material is very low, the
large reconnection rate required to power a flare demands the presence of 
reconnecting structures with very small length scales -- namely, current 
sheets.
Both in solar and magnetospheric physics, the study of particle
acceleration in such structures has received considerable attention
over the past few decades. Under non-steady state conditions, a reconnecting 
magnetic field produces an inductive electric field $\mathbf{E}$: the 
resistive magnetohydrodynamic (MHD) form of Ohm's law indicates that in 
general this $\mathbf{E}$ has a component parallel to the local magnetic 
field, and can thus accelerate charged particles. An example of such a 
non-steady state is a perturbed reconnecting X-type neutral point, for which 
Craig and McClymont (1991) calculated the normal modes in a strictly 2-D 
field, ${\bf B}(x,y)$. In this paper we demonstrate that the analysis of Craig
and McClymont is applicable when a $z$-invariant component of the magnetic 
field is included, enabling it to be extended to configurations more likely to 
be representative of conditions in the flaring solar corona.

In the context of solar flare physics, a current sheet is often
envisaged as arising in a two-ribbon flare, where reconnection of
oppositely-directed, predominantly vertical magnetic fields results in
a sheet structure having a length (parallel to the solar surface) on
the order of the length of a post-flare arcade ($\sim 10^7$m), a
vertical extent (height perpendicular to the solar surface)
comparable to this, and a thickness on the order of a few times the
ion gyroradius. Particle acceleration - in particular proton
acceleration - in just such a (collisionless) reconnecting current
sheet was addressed by Martens (1998). Following from the magnetotail
work of Speiser (1965) he considered the effects of the inclusion of a
component of magnetic field perpendicular to the plane of the current
sheet, deriving expressions for the maximum energy attained before the
gyromotion around this perpendicular field causes the protons to exit
the sheet, and hence the acceleration region. With reconnection
electric field strengths comparable with those observed (e.g. Kopp and
Poletto, 1986), it was found that the proton energy and flux budgets
for flares were compatible with those arising from acceleration in a
macroscopic current sheet (for example formed in the wake of a rising
filament, with vertical extent on the order of $10^7$m, comparable to
its length). 

Litvinenko and Somov (1993) also considered the current sheet
geometry, demonstrating the effects of the addition of a further
magnetic field component parallel to the direction of the reconnection
electric field. A component of the charged particles' motion is then
gyration about this parallel field, which allows them to stay longer
in the current sheet, and attain higher energies. It is this fact
which also becomes important in our studies.

While a reconnecting current sheet with finite (and large) vertical
extent is frequently used in considering acceleration in large
two-ribbon flares, where such a configuration is expected on the basis
of numerical simulations of filament lift-off, we concern ourselves here
with a current sheet of zero vertical extent, i.e. a coronal X--line,
which corresponds to the initial stages of reconnection behind a
filament. In the simplest case, this configuration is identical in 
cross-section [the $(x,y)$ plane] to a 2--D X--type neutral point, but is 
invariant in the $z$--direction. In reality, true magnetic nulls 
($B \to 0$) are likely to be rare in the corona. The 
elaboration of this scenario which we study here, and find to have important 
consequences for particle acceleration, is the inclusion of a component of 
magnetic field also in the $z$--direction.

Proton acceleration at an X--point with finite $B_z$ was investigated by 
Bulanov (1980), Mori, Sakai, and Zhao (1998), Browning and Vekstein (2001), 
and Bruhwiler and Zweibel (1992). The present study differs from that of 
previous workers in 
that the accelerating electric field $E_z$ is obtained self--consistently from
a magnetic flux function corresponding to a reconnecting eigenmode of the 
X--point. This perturbation solution was invoked in the context of particle 
acceleration also by Petkaki and MacKinnon (1997), but without the addition of
the finite $B_z$ component.

A major challenge for any flare acceleration mechanism is to obtain
sufficient high energy particles in a short timescale; new
observations from the RHESSI satellite are putting ever tighter
constraints on timescales, spectra and total energies. The typical
requirements for protons, as summarised by Miller (1998)
are as follows: they are accelerated up to energies of $\sim$ 100~MeV
on timescales of about a second, and to about a GeV on timescales of a
few seconds. Proton acceleration lasts for several tens of seconds, at
a rate of $\sim 10^{35}{\rm s}^{-1}$, such that the total energy
content in protons above an MeV is $\sim 10^{24}$ J. Given a coronal
density of $10^{15}-10^{16}{\rm m^{-3}}$ and volume of $\sim
10^{21}{\rm m^{-3}}$, this implies that each second around
1-10\% of all coronal protons must be accelerated to MeV energies (and
therefore require rapid replenishing - we do not address this
here). We find that the inclusion of a moderate longitudinal field
component greatly assists in this process.

This paper is structured in the following way. In
Section 2 we describe our new algorithm for particle
calculations. Section 3 describes the model employed in the 
simulations. Some sample simulation results and the effect of varying 
parameters of the simulation, such as the strength of the $z$ component of the
magnetic field, are described in Section 4 and we 
end with discussions and conclusions in Section 5.

\section{Energy--Conserving Algorithm} 

The nonrelativistic Lorentz force equations
$$m\frac{d\mathbf{v}}{dt} = Ze\mathbf{v} \times
\mathbf{B}(\mathbf{x})+Ze\mathbf{E}(\mathbf{x}), \;\;\; \frac{d\mathbf{x}} 
{dt} = \mathbf{v}, \eqno (1) $$
are approximated in the CUEBIT (CUlham Energy--conserving orBIT) code by the 
following finite difference equations
$$m\frac{\mathbf{v}^{i+1} - \mathbf{v}^i}{\Delta t}
=Ze\left(\frac{\mathbf{v}^{i+1}+\mathbf{v}^i}{2}\right)  
\times\mathbf{B}\left(\frac{\mathbf{x}^{i+1} +
\mathbf{x}^i}{2}\right)
+ Ze\mathbf{E}\left(\frac{\mathbf{x}^{i+1} + \mathbf{x}^{i}}{2}\right), \eqno 
(2) $$
$$\frac{\mathbf{x}^{i+1} - \mathbf{x}^i}{\Delta t} =
\frac{\mathbf{v}^{i+1} + \mathbf{v}^i}{2}. \eqno (3) $$
Here $m$, $Ze$ denote particle mass and charge. 
At the start of each timestep $\mathbf{x}^{i+1}$ is set equal to 
$\mathbf{x}^{i}$ in a first calculation of $\mathbf{v}^{i+1}$. A few 
iterations are made which converge quickly to a final $\mathbf{v}^{i+1}$. 

Mori, Sakai and Zhao (1998) also used a scheme in which ${\bf v}$ was set 
equal to $({\bf v}^{i+1}+{\bf v}^i)/2$ on the right hand side of the
Lortentz force equation. In the special case where $\mathbf{E}=0$, the scalar 
product of the right hand side of Equation (2) with 
$\mathbf{v}^{i+1}+\mathbf{v}^i$ is identically zero, so that 
$(v^{i+1})^2=(v^i)^2$. The 
scheme thus conserves energy exactly. This makes it possible 
to obtain accurate results with relatively long timesteps. A modified version 
of Equation (2) conserves total energy exactly when 
$\mathbf{E}$ is a finite potential electric field. For non-potential electric 
fields, such as those arising from magnetic reconnection, we have found the 
method remains accurate for large timesteps (exceeding the Larmor period in 
most of the computational domain), allowing large numbers of particles to be 
simulated for timescales that are relevant for flare acceleration. CUEBIT
has also been benchmarked by using it to compute energetic particle orbits 
in magnetic fusion experiments (Wilson et al., 2002), and will in the future 
be used to study particle transport under various conditions (e.g. in the 
presence of turbulent electromagnetic fields) in such experiments. 

For the simulations discussed in Section 4 it is 
not necessary to incorporate relativistic kinematics in Equation 
(2) as the energies reached are only a small fraction
of the particles' rest mass energy (we consider only proton acceleration). 
However, a relativistic version of the code is currently being developed
for the purpose of describing electron acceleration and transport.  

\section{Model of a reconnecting X-type structure}

\subsection{Craig and McClymont solution}

For a simple model of a reconnecting field at an X-type neutral point
we use the two--dimensional description of Craig and McClymont (1991), with 
the additional element of a finite third magnetic field component $B_z$.
To determine the conditions under which the Craig and McClymont analysis 
applies with $B_z \ne 0$, we re-derive below their equations for the 
evolution of a flux function $\psi$, defined such that the curl of 
$\mathbf{A}\equiv\psi\hat\mathbf{z}$ is equal to the magnetic field in 
the $(x, y$) plane. The induction equation
$$\frac{\partial \mathbf{B}}{\partial t} = \nabla\times
\left(\mathbf{v}\times\mathbf{B}\right)+\frac{\eta}{\mu_{0}}
\nabla^{2} \mathbf{B}, \eqno (4) $$
can be written in the form
$$\frac{\partial \mathbf{A}}{\partial t} = \mathbf{v} \times \left(
\nabla \times\mathbf{A} \right) + \frac{\eta}{\mu_{0}}\nabla^{2} \mathbf{A}.
\eqno (5) $$
Here $\mathbf{v}$ is flow velocity, $\eta$ is resistivity (assumed to be 
constant) and $\mu_0$ is the permeability of free space. The first term on the
right hand side of Equation (5) can be expanded to 
give
$$\mathbf{v}\times\left(\nabla\times\mathbf{A}\right) = \nabla
\left(\mathbf{v}\cdot\mathbf{A}\right)-\left(\mathbf{v} \cdot
\nabla\right)\mathbf{A} - \mathbf{A} \times \left(\nabla
  \times \mathbf{v}\right) - \left(\mathbf{A} \cdot
  \nabla\right)\mathbf{v}. \eqno (6) $$
We assume that flows only occur in the $(x, y)$ plane and that there are no 
variations in the $z$-direction. In these circumstances the first, third and 
fourth terms on the right hand side of Equation (6) are all zero and Equation 
(5) reduces to   
$$\frac{\partial\psi}{\partial t} + \left(\mathbf{v}\cdot\nabla\right)\psi= 
\frac{\eta}{\mu_{0}}\nabla^{2}\psi, \eqno (7) $$
which is the form of the induction equation used by Craig and McClymont.

Neglecting plasma pressure, viscosity 
and any external forces such as gravity
the momentum equation is
$$\frac{\partial\mathbf{v}}{\partial t}+(\mathbf{v}\cdot\nabla)\mathbf{v} = 
\frac{1}{\rho}\mathbf{j}\times\mathbf{B}, \eqno (8) $$
where $\mathbf{j}$, $\rho$ denote current and mass density. Now in the present
case
$$\mathbf{j}=\frac{1}{\mu_0}\nabla\times\mathbf{B} = -\frac{\hat\mathbf{z}}
{\mu_0}\nabla^2\psi. \eqno (9) $$
Substituting this expression into Equation (8) and noting that 
$\nabla\psi$ is in the $(x,y)$ plane, we obtain
$$\frac{\partial\mathbf{v}}{\partial t}+(\mathbf{v}\cdot\nabla)\mathbf{v}=
-\frac{1}{\mu_0\rho}\left(\nabla^2\psi\right)\nabla\psi, \eqno (10) $$
i.e. Equation (2.3) in Craig and MyClymont (1991). 
 
\subsection{Extension of Craig and McClymont solution to $B_z \ne 0$}
 
We now assume that the $\mathbf{B}$ field is of the form
$$\mathbf{B} = \nabla\times(\psi\hat\mathbf{z}) + B_z\hat{\mathbf{z}}, \eqno 
(11) $$
where $B_z$ is constant and uniform. As before, we assume that there are
no variations in the $z$ direction. Substituting Equation 
(11) into Equation (4) and writing
$\mathbf{A}=\psi\hat\mathbf{z}$ as before gives
$$\frac{\partial}{\partial t}\left(\nabla\times\mathbf{A}+B_z\hat{\mathbf{z}}
\right) = \nabla\times\left(\mathbf{v}\times\left(\nabla\times\mathbf{A}+B_z
\hat{\mathbf{z}}\right)\right)+\frac{\eta}{\mu_0}\nabla^2
\left(\nabla\times\mathbf{A}+B_z\hat{\mathbf{z}}\right). \eqno (12) $$
Since $B_z$ is constant and uniform this reduces to
$$\frac{\partial}{\partial t}\left(\nabla\times\mathbf{A}\right)=\nabla\times
\left(\mathbf{v}\times\left(\nabla\times\mathbf{A}\right)\right)+\nabla\times
\left[\mathbf{v}\times\left(B_z\hat{\mathbf{z}}\right)\right]+
\frac{\eta}{\mu_{0}}\nabla^2\left(\nabla\times\mathbf{A}\right). \eqno (13) $$
This is equivalent to the induction equation of Craig and McClymont except 
for the term containing $B_z$ on the right hand side. This term can be
written in the form
$$\nabla\times\left[\mathbf{v}\times\left(B_z\hat{\mathbf{z}}\right)\right]=-
\left( B_{z} \hat{\mathbf{z}} \right)\nabla \cdot \mathbf{v}. \eqno (14) $$
Since $B_z$ is not necessarily zero, the only way for this quantity
to be zero is if $\nabla\cdot\mathbf{v} = 0$, i.e. the plasma flow must
be incompressible. This assumption is often invoked in studies of magnetic 
reconnection (e.g. Biskamp, 2000; Priest and Forbes, 2001), and is a 
reasonable one for solar flare plasmas. Incompressibility 
is implicit in the model of Craig and 
McClymont: they use Equation (10) in a dimensionless form 
that is only valid if the plasma density is constrained to be 
uniform and time--independent. In the limit of ideal MHD, it is 
straightforward to show that the flow must be incompressible when, as in the 
scenario considered here, there are no variations in the $z$--direction and 
$B_z$ is finite (Strauss, 1976). With the assumption of incompressibility,
Equation (7) is valid for finite $B_z$. Moreover, 
since $B_z$ is curl-free it does not contribute to the current; since the 
latter is oriented in the $z$ direction 
[Equation (9)], it follows that $B_z$ does not contribute to 
the Lorentz force either, and thus the momentum equation of Craig and 
McClymont [Equation (10)] also remains valid. The 
solutions of these equations derived by Craig and McClymont are therefore 
applicable when the magnetic field has a uniform and static $z$ component.

\subsection{Perturbation solution}

The equilibrium X--type neutral point structure invoked by Craig and McClymont
(1991) is described by
$$\mathbf{B} = \frac{B_{o}}{R_{o}} \left(y \hat{\mathbf{x}} + x
\hat{\mathbf{y}} \right), \eqno (15) $$
where $B_o$ is the field strength at the (circular) boundary
of the system where $R=R_{o}$,
$R$ being the radial co-ordinate measured from the z-axis outwards.
Perturbations to this field
geometry will result in normal modes of oscillation. Following 
Petkaki and MacKinnon (1997), we consider only the fundamental mode, with zero
azimuthal and radial mode numbers. Approximate analytical solutions for the 
perturbed magnetic and electric fields can be obtained by dividing the
system into an ideal outer region and a resistive inner region, separated by 
a critical radius
$$ R_{c} = R_{o} \left( \frac{2}{S} \right)^{\frac{1}{2}}, \eqno (16) $$
where $S$, the Lundquist number at the boundary $R=R_o$, is equal to 
$\mu_0R_ov_{A}/\eta$ where $v_A$ is the Alfv\'en speed at $R = R_{o}$ . The 
solution for the perturbed magnetic field in cylindrical polar coordinates can
be approximated by (Craig and McClymont, 1991)
$$\delta B = -dB\frac{R_{o}}{R}\omega\cos
\left(\omega\ln\frac{R}{R_o}\right)\cos\left(\frac{\omega
v_A}{R_o}t\right)\exp\left(-\frac{\alpha v_A}{R_o}t\right)
\hat{\mathbf{\varphi}}, \eqno (17) $$
for $R > R_{c}$. Here $dB$ is an arbitrary scaling parameter, $\omega$ is
a dimensionless mode frequency, $\alpha$ is a dimensionless decay constant and 
$\hat{\varphi}$ is the unit vector in the azimuthal direction.
For $R \le R_{c}$ the magnetic field perturbation is zero. The value of 
$\omega$ is determined by the requirement that Equation 
(17) yields $\delta B = 0$ at $R = R_c$, and $\alpha
\simeq \omega^2/2$ (Craig and McClymont, 1991). 
The electric field resulting from this solution is wholly in
the $z$--direction and is given by
$$ E_z = dB v_A \sin \left( \omega \ln \frac{R}{R_{o}}\right)\exp\left(-
\frac{\alpha v_A}{R_{o}}t\right)\left(\alpha\cos\left(\frac{\omega v_{A}}
{R_o}t\right)+\omega\sin\left(\frac{\omega v_{A}}{R_{o}} t \right)\right),
\eqno (18) $$
for $R > R_c$ and
$$E_{z} = -dB v_A \exp \left( - \frac{\alpha v_A}{R_{o}} t
\right)\left( \alpha \cos \left( \frac{\omega v_{A}}{R_{o}} t
  \right) + \omega \sin \left( \frac{\omega v_{A}}{R_{o}} t \right)
\right), \eqno (19) $$
for $R \le R_{c}$. Equations (17), (18), 
and (19) remain valid for an incompressible plasma with
finite $B_z$ constant in time and space. We show below that the 
inclusion of a modest $B_z$ dramatically increases the efficiency with which
test particles are accelerated in this magnetic geometry.

\section{Acceleration in a reconnecting X-type structure}

\subsection{Choice of parameters}

We consider specifically the acceleration of test particle protons in
a prescribed field of the type discussed in the previous section. It
is known that protons are accelerated in flares to energies of at
least several tens of MeV in a timescale of the order of one second
(Miller, 1998; Aschwanden, 2002). A straightforward
calculation indicates that this could be achieved with a parallel
electric field $E_{\parallel}$ of around 1$\,$Vm$^{-1}$. Such field
strengths are, incidentally, consistent with the parallel electric
fields implied by observations of the separation rates of flare
ribbons, e.g. Kopp and Poletto (1986). 
However, as noted recently by Craig
and Litvinenko (2002), fields of this magnitude
are not consistent with classical Spitzer resistivity.
In deriving their reconnecting field solutions, Craig and McClymont
(1991) used the resistive MHD form of Ohm's law, i.e.
$E_{\parallel}=\eta j_{\parallel}$ where $j_{\parallel}$ is the
parallel component of ${\bf j}$.
The maximum possible current density in a hydrogen plasma is $j_{\rm
max}=2nec$, where $n$ is particle density and $c$ is the speed of
light. So a coronal density of $10^{15} - 10^{16}\ {\rm m}^{-3}$
implies that $j_{\rm max} \sim 10^5 -10^6\,$Am$^{-2}$. However, if the
resistivity were determined by classical electron--ion collisions, and
the plasma temperature $T$ were equal to $10^7\,$K, a field of
1$\,$Vm$^{-1}$ would produce a steady--state current density of around
$2\times 10^7\,$Am$^{-2}$. This high current cannot be sustained by a
classical collisional resistivity. We deduce that protons can only be
accelerated to $\gamma$--ray--emitting energies in a reconnecting
field of the type invoked by Craig and McClymont if $\eta$ is much
higher than the Spitzer value, i.e. it is anomalous. Craig and Litvinenko
(2002) have pointed out another reason for ruling out Spitzer resistivity in 
the flaring corona, namely that it implies a resistive scale length that is 
smaller than the collisional mean free path.

Anomalous resistivity can result from lower hybrid or ion acoustic 
micro--turbulence. Whether lower hybrid or ion acoustic waves are more 
likely to produce the required anomalous resistivity depends on several 
parameters, including the ratio of plasma pressure to magnetic field pressure,
the electron to ion temperature ratio, and the direction of the current 
relative to the local magnetic field (Aparicio et al., 1998; Biskamp, 2000).
In either case, turbulence results from the 
current density exceeding a certain threshold $\kappa nec_s$ where $c_s$ is 
the sound speed and $\kappa$ is a numerical factor typically ranging from 
unity up to about the square root of the ion to electron mass ratio (Kulsrud, 
1998; Aparicio, {\it et al.} 1998). The anomalous resistivity then prevents
the current density from exceeding this threshold: the turbulence
increases the effective collisionality of the plasma. In the case of
ion acoustic turbulence, for example, there is a strong interaction
with protons whose velocity component $v$ parallel to the wave
propagation direction is of the order of $c_s$
(Ishihara and Hirose, 1981). The strength of the interaction falls off
roughly as $1/v^2$, and is negligibly small for particles lying
sufficiently far out in the tail of a Maxwellian distribution
(cf. Litvinenko and Somov, 1993). Therefore, some small fraction of 
the initial proton population is effectively collisionless, and can
be described using Equation (1): it is this
sub-population of protons which we simulate using CUEBIT.
The bulk of the plasma is effectively collisional, due to the 
anomalous resistivity. The use of a test particle approach requires that
any net current associated with accelerated particles is small compared to 
the current corresponding to the reconnecting field.

For simplicity, the particles in most of our simulations were given zero 
initial velocity. This is permissible, since the thermal spread of
velocities corresponding to a typical coronal temperature is very
small compared to the near--relativistic speeds required for protons
to excite $\gamma$--ray emission, and we will show that that the final
proton energy spectrum above $\sim 10\,$keV in a typical simulation does not 
change significantly when the $\delta$--function initial velocity
distribution is replaced with a $10^7$K Maxwellian.

The choice of parameters in the simulations is dictated in part by 
the limits imposed by anomalous resistivity, combined with the typical
parallel field strengths needed. We assume, for definiteness 
$n=10^{16}$m$^{-3}$, $R_o=10^7$m and $B_o=0.01\,$T. If the temperature is 
assumed to be 10$^7$K, and the current is assumed to be limited to $nec_s$
(i.e. $\kappa=1$), the anomalous resistivity required for 
$E_{\parallel}=1\,$Vm$^{-1}$ is then $1.4 \times 10^{-3}\,\Omega$m and the 
Lundquist number at the boundary $R=R_o$ (a key parameter in the Craig and 
McClymont model) is $S \simeq 2\times 10^{10}$. Equations 
(3.6) and (3.8) of Craig and McClymont (1991) then yield $\omega \simeq 0.13$ 
and $\alpha \simeq 0.0088$. The scaling parameter $dB$ in Equations 
(17), (18) and (19) was  
chosen to give $E_z\simeq1\,$Vm$^{-1}$ at $t=0$ and $R < R_c$ (in this region 
$E_z \simeq E_{\parallel}$) rising to approximately 1.4\,Vm$^-1$ after 1$\,$s.
For comparison, simulations were also carried 
out with $E_z=0.1\,$Vm$^{-1}$ and $10\,$Vm$^{-1}$. Strictly speaking, a change
in $E_z$ implies a change in $\eta$ and hence $S$ if the limiting current is 
assumed to be a fixed multiple of $nec_s$. However, since $\omega$ and 
$\alpha$ have only a weak (logarithmic) dependence on $S$, the same values
of these parameters were used in all the simulations.

The only remaining parameter to fix is $B_z$. It follows from the conclusions
drawn in Section 3.2 that this can be varied freely 
without any of the other parameters being affected. Simulations were carried
out with $B_z=0\,$T, $10^{-5}\,$T, $10^{-4}\,$T, $10^{-3}\,$T and $10^{-2}\,$T:
these are all reasonable values for active region coronal magnetic fields.
       
\subsection{Results}
  
In each simulation the trajectories were computed of approximately $10^6$ 
protons with an initially uniform random distribution of positions in the 
reconnection region $\left(R\le R_{o}\right)$, and, unless otherwise stated, a
$\delta$--function velocity distribution. The duration of each 
simulation, 1$\,$s, was chosen on the basis that protons appear to be 
accelerated to tens of MeV on timescales of this order 
(Miller, 1998; Aschwanden, 2002). The timestep $\Delta t$ was of the
order of 100 Larmor periods calculated at $R=R_{o}$ (simulations with
shorter timesteps produced essentially identical results); there were
approximately 1500 timesteps per simulation. The value of $dB$ was
chosen such that the magnitude of the magnetic field perturbation at
$R=R_{o}$ and $t=0$ was always $\le 10^{-4}$~T.

Figure 1 shows $R$ and particle energy $E$ as 
functions of time for a proton initially lying close to the X-point ($R=10$m)
in field configurations with $B_z=0$ (solid curves) and $B_z=10^{-4}\,$T 
(dashed curves). The other field parameters are those listed at the end of
the previous section. The proton initially lies at azimuthal
angle $\varphi=24^{\circ}$ in the ($x,y$) plane, and has velocity components
in the cylindrical coordinate system $v_R=v_{\varphi}=v_z=10^5\,$ms$^{-1}$.

It is immediately clear that the addition of even a very modest $B_z$ has a
dramatic effect on the trajectory of the particle in phase space. When $B_z=0$
the total magnetic field is very small close to the X-point (since $\delta 
B=0$ in this region). The Lorentz force in the ($x,y$) plane is consequently 
very weak, the particle is effectively unmagnetised, and hence moves rapidly  
away from the X-point. There is only a short phase of 
acceleration (in this case to about 1~keV): far from the X--point, the 
particle becomes magnetized and, since ${\bf B}$ is strictly perpendicular to 
${\bf E}$, there is no further acceleration. However, a longitudinal field of 
$10^{-4}$T is sufficient to confine the particle to within a few tens of 
metres of the X-point, with the result that acceleration to energies of 
100~keV and above can easily occur. Because $B_z$ is assumed to be uniform 
and the field components in the ($x,y$) plane are very small, the combined 
effect of grad--$B$, curvature and ${\bf E}\times{\bf B}$ drifts on the 
particle trajectory is negligible.     

\begin{figure}[htb]
\setlength{\unitlength}{1cm}
\begin{picture}(12.0,6.0)        
\put(0.0,-11.5){\includegraphics{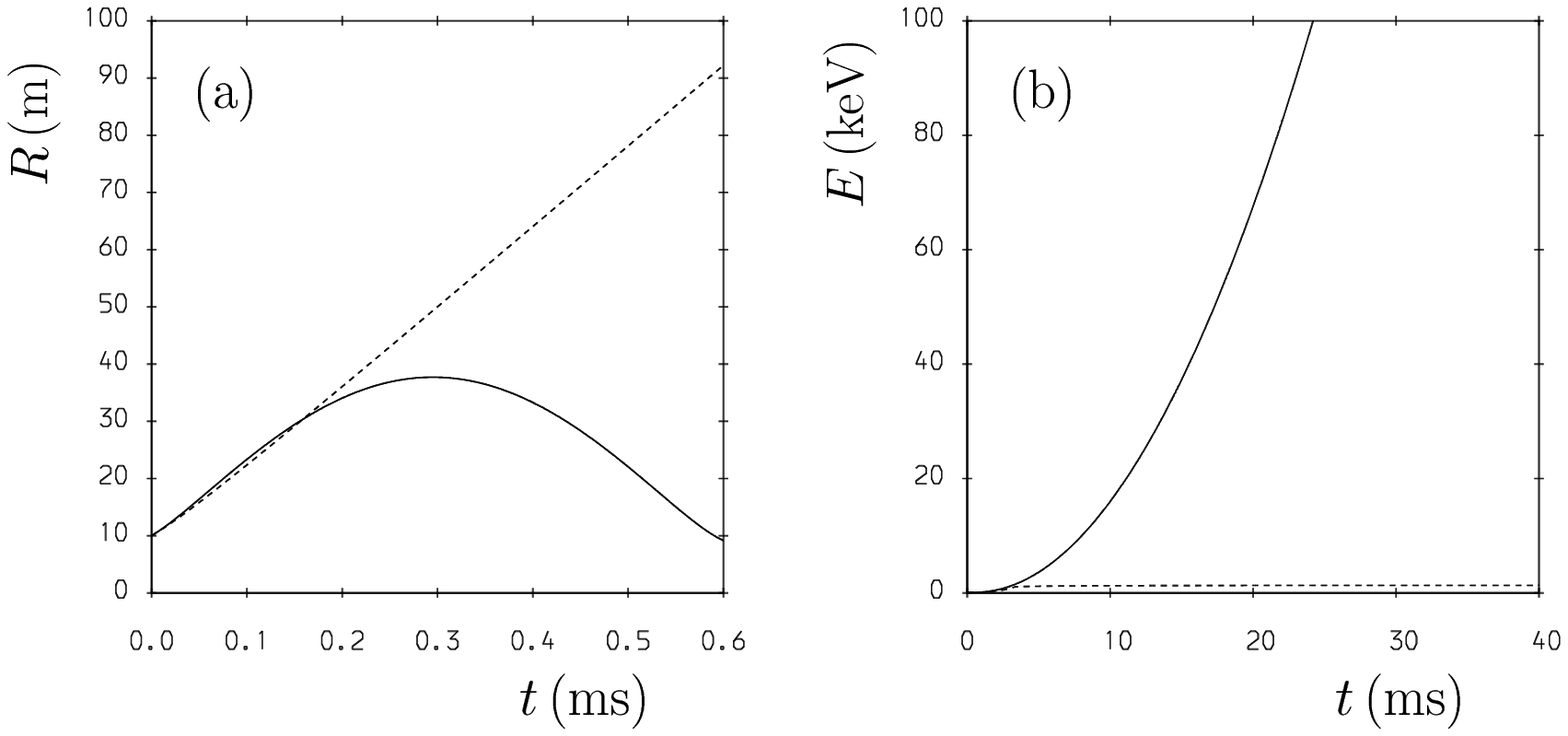}}
\end{picture}
\end{figure}

\rif{{\bf Fig. 1.} Distance from the X--point versus time 
for particles initially lying close to $R=0$ in field configurations 
with $B_z=0$ (solid curve) and $B_z=10^{-4}\,$T (dashed curve). (b)
Energy versus time over a longer timescale for the particles whose 
drifts are shown in (a).}

\vskip 0.5cm

Figure 2 shows final proton energy spectra for $B_z=10^{-4}$T and 
initial electric field in the inner resistive region $E_{0z}=1$Vm$^{-1}$.
Two different initial proton distributions were used: a $\delta$-function 
(faint curve) and a $10^7$K Maxwellian (bold curve). In both cases, protons 
are accelerated to energies of up to several MeV. The only significant
differences between the two distributions occur at energies $< 10\,$keV:
above this energy, the final distribution is insensitive to the initial
conditions. 

\begin{figure}[htb]
\setlength{\unitlength}{1cm}
\begin{picture}(8.0,8.0)(0.0,0.0)        
\put(2.5,0.0){\includegraphics{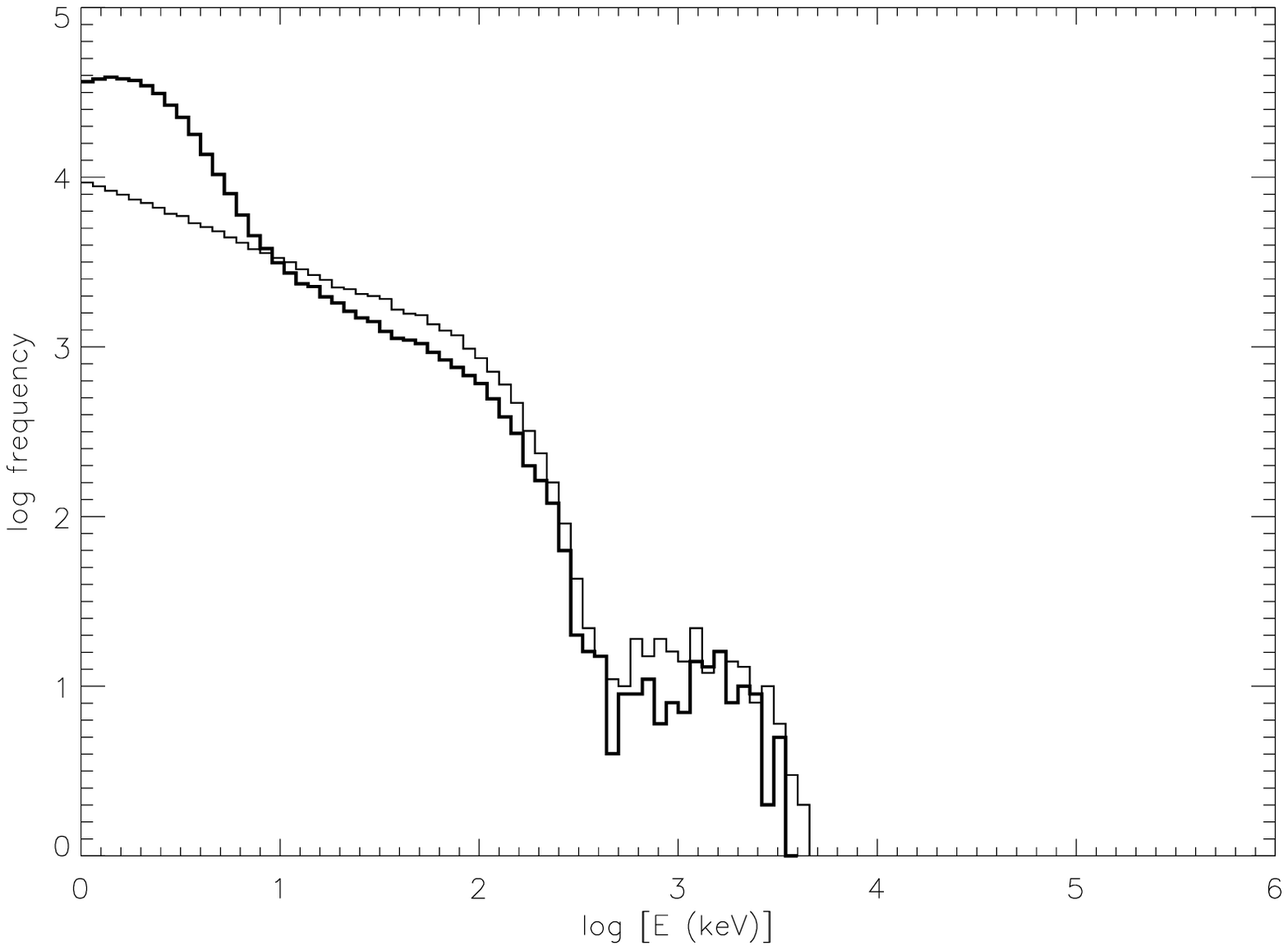}}
\put(4.0,5.1){$T=0$}
\put(4.7,6.3){$T=10^7$K}
\end{picture}
\end{figure}

\rif{{\bf Fig. 2.} Final proton energy
distributions for $B_z=10^{-4}\,$T, $E_{0z}=1$Vm$^{-1}$. The initial 
distributions have temperature $T=0$ (faint curve) and $T=10^7$K
(bold curve).}

\vskip 0.5cm

Figures 3 and 4 show final proton energy distributions for several pairs of
values of $B_z$ and $E_{0z}$. In every case a $\delta$-function initial 
distribution was used.  
In Figure 3 $B_z=10^{-4}$T and $E_{0z}$ ranges from 0.1~Vm$^{-1}$ to 
10~Vm$^{-1}$. Not surprisingly, protons are accelerated to higher energies 
as $E_{0z}$ is increased. In contrast to the results obtained by Mori, Sakai, 
and Zhao (1998), the shape of the accelerated proton spectrum depends strongly
on the model parameters: although in some cases the protons have a power law
spectrum at low energy (1--100~keV for $E_{0z}=1$~Vm$^{-1}$), the
power law index varies considerably. It should be noted that the frequencies 
$f$ plotted in Figures 3 and 4 represent the number of particles at the end of
each simulation with energies in a fixed range of values of log$_{10}E$, i.e. 
$$ f(E)d({\rm log}_{10}E) \propto F(E)dE, \eqno (20) $$
where $F$ is the true energy distribution. Thus, $F(E) \propto f(E)/E$. 
At low energy, the distributions in Figure 3 correspond to $F(E) \propto 
E^{-\gamma}$ where $\gamma \le 2$: Mori, Sakai, and Zhao (1998) obtained 
$\gamma \simeq 2.0-2.2$.       
      
Another striking feature of the distributions in Figure 3 is the formation of a
bump-on-tail at high energy. This also occurs in the simulations of Petkaki 
and MacKinnon (1997), but not in 
those of Mori, Sakai and Zhao (1998). In general, models in which there is 
preferential acceleration of a small sub-population of protons to MeV energies
and above are more efficient than models which predict monotonic decreasing 
spectra, since it is only at high energy that direct observational evidence
for accelerated protons exists. The 
occurrence of bumps-on-tail in our simulations and those of Petkaki 
and MacKinnon, and their absence from those of Mori, Sakai and Zhao 
(1998), appears to be due to the choice of field configuration used in 
these studies. Mori and co-workers assumed a purely hyperbolic magnetic field
in the ($x,y$) plane (i.e. $\delta B = 0$) and a uniform $E_z$. With finite 
$B_z$, there is then a large parallel electric field component $E_{\parallel}=
E_zB_z/B^2$ throughout the computational domain, and all the 
test particles are susceptible to strong acceleration. In our case, 
$E_{\parallel}$ is significantly reduced outside the critical radius 
$R=R_c$, due to the presence of a perturbation to the hyperbolic magnetic
field ($\delta B \ne 0$) and also a fall--off in $E_z$ associated with   
the spatial profile of the reconnecting mode eigenfunction [cf. Equation 
(18)]. Consequently, particles initially lying inside $R=R_c$ are subject to 
stronger acceleration than those initially lying outside.   

\begin{figure}[htb]
\setlength{\unitlength}{1cm}
\begin{picture}(8.0,8.0)(0.0,0.0)        
\put(2.5,0.0){\includegraphics{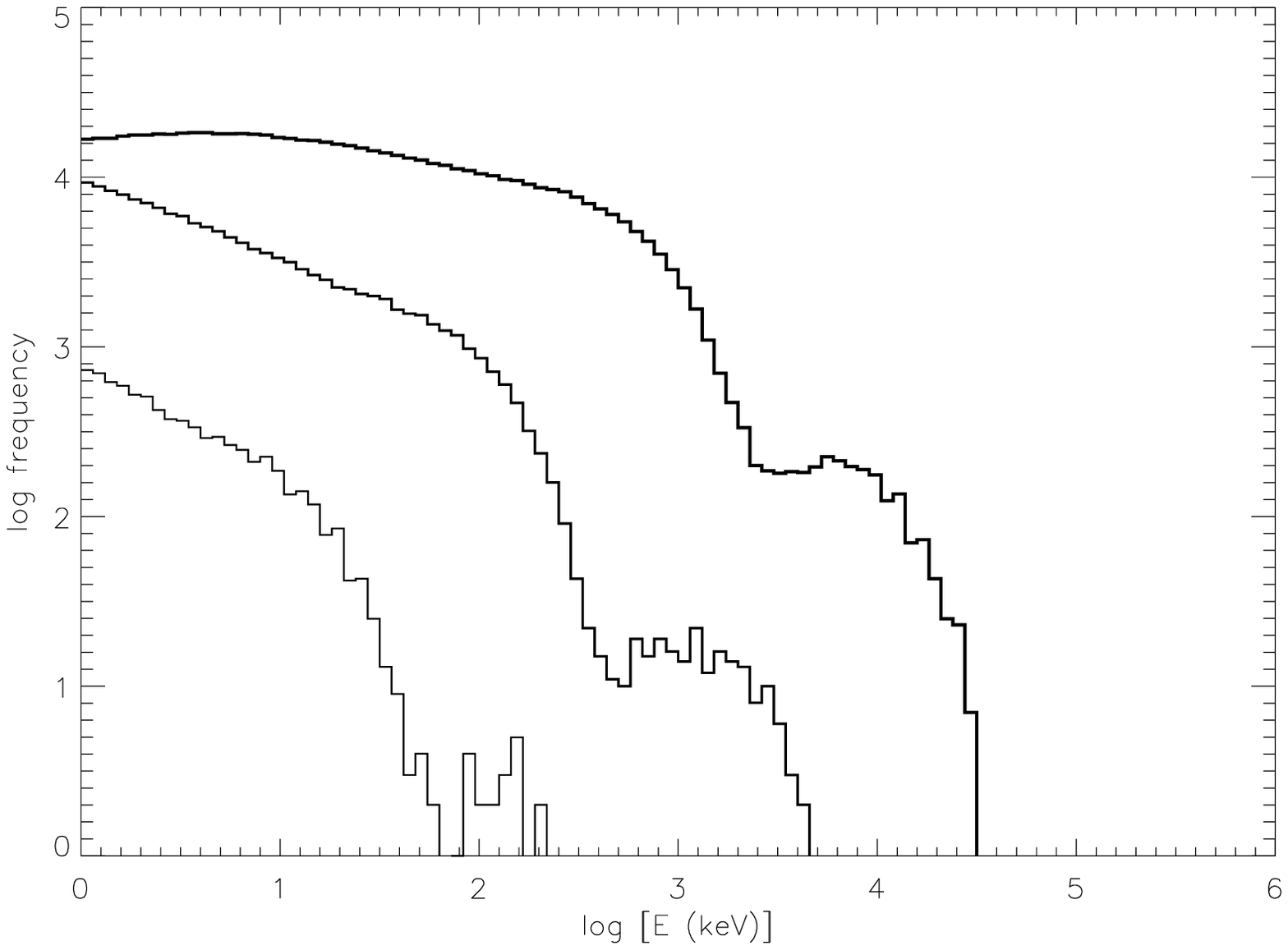}}
\put(4.3,2.3){$0.1\,$Vm$^{-1}$}
\put(6.1,3.2){$1\,$Vm$^{-1}$}
\put(8.6,4.5){$E_{0z} = 10\,$Vm$^{-1}$}
\end{picture}
\end{figure}

\rif{{\bf Fig. 3.} Final proton energy
distributions for $B_z=10^{-4}\,$T and three values of $E_{0z}$.}

\vskip 0.5cm

Final proton distributions are shown in Figure 4 for $E_{0z}=1~$Vm$^{-1}$ and 
four different values of $B_z$. Results are not shown for $B_z=0$ since in 
this case the number of particles accelerated to energies of more than 1~keV 
was negligible. It can be seen that protons are accelerated to progressively
higher energies as $B_z$ is increased -- in the case of $B_z=10^{-2}$T, 
particle energies of up to 40~MeV are observed. This appears to be due simply 
to improved particle confinement close to the X-point: the amplitude of the 
sinusoidal variation in Figure 1(a) varies as $1/B_z$ and the drift speed
varies as $1/B_z^2$.  For $B_z < 10^{-2}$T bump-on-tail formation at high 
energy is again apparent. 

\begin{figure}[htb]
\setlength{\unitlength}{1cm}
\begin{picture}(8.0,8.0)(0.0,0.0)        
\put(2.5,0.0){\includegraphics{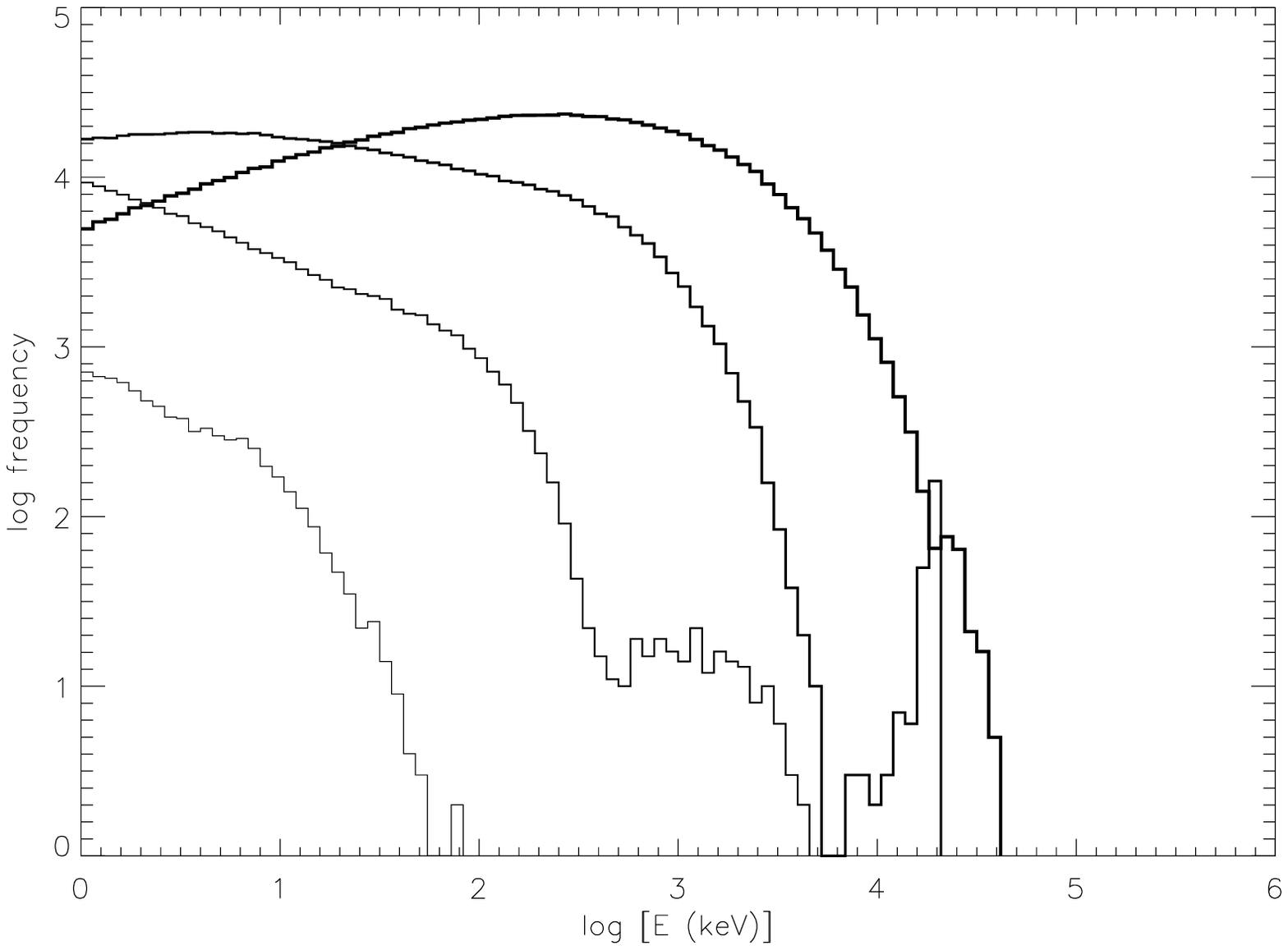}}
\put(5.0,1.8){$10^{-5}\,$T}
\put(6.2,2.8){$10^{-4}\,$T}
\put(7.5,4.0){$10^{-3}\,$T}
\put(10.0,4.5){$B_{z} = 10^{-2}\,$T}
\end{picture}
\end{figure}

\rif{{\bf Fig. 4.} Final proton energy distributions for 
$E_{0z}=1\,$Vm$^{-1}$ and four values of $B_z$.}

\vskip 0.5cm

\section{Discussion and Conclusions} 

The addition of a longitudinal field component to a two--dimensional 
reconnecting magnetic configuration massively increases the efficiency of 
particle acceleration in such configurations. Physically, this is due to the 
fact that particles close to the magnetic X-point are strongly magnetized by 
the longitudinal field and are not subject to strong grad--$B$, curvature 
or ${\bf E}\times{\bf B}$ drifts. The effect of a small but finite 
longitudinal field $B_z$ on particle acceleration is so dramatic that $B_z=0$ 
should be regarded as a singular case, unlikely to be representative of the
conditions prevailing in the flaring corona. MeV protons can still be produced
when $B_z=0$, but the number of particles accelerated to such energies is only
significant if the resistivity $\eta$ is assumed to be extremely
large -- much larger even than the anomalous values corresponding to ion 
acoustic or lower hybrid turbulence (Petkaki and MacKinnon, 1997). A key 
result of the present study is that the presence of a longitudinal guide field
makes it possible for protons to reach MeV energies in a plasma with 
realistic values of $\eta$.        

In the case of the simulation with $B_z=10^{-2}$T, $E_{0z}=1~$Vm$^{-1}$, 14\% 
of the protons reach energies exceeding 1 MeV in 1s. However, as discussed in 
Section 4.1, only a fraction of the initial particle population is 
modelled using the collisionless equations of motion, and so the simulation
results imply a super-MeV proton fraction of much less than 14\%. As noted 
in Section 1, observations imply that up to 1-10\% of the protons in the 
flaring corona are accelerated per second. While the test particle method
adopted here yields valuable insights into the physics of proton 
acceleration in flares, a more self-consistent approach may be required  
to meet the tight constraints on energetic proton fluxes imposed by recent 
observations. 

The sensitivity of the final distributions to $B_z$ (Figure 4) suggests that 
it might be possible to set constraints on the value of this parameter using 
$\gamma$--ray observations. However, it is not clear to what extent the
simulation results depend on other features of the model, such as the 
assumption of an azimuthally-symmetric, centrally-peaked  perturbation to 
the magnetic flux. CUEBIT could be used to study particle acceleration in 
X-point field configurations with a range of different normal mode 
perturbations. We also intend to study electron acceleration: the numerical 
scheme in Equations (2) and (3) allows 
sufficiently large timesteps (relative to the Larmor period) to make this 
feasible, although it will probably be essential to use a fully relativistic 
version of the code in this case. Collisions can also be added to the scheme 
in a straightforward way, enabling simulations to be carried out on longer 
timescales.

In analytical studies of particle acceleration at reconnecting 
current sheets Litvinenko and Somov (1993) and Litvinenko (1996) 
formulated expressions for critical values of $B_z$, such that particles are 
efficiently accelerated, and for the typical energies which they can reach. 
Our results are not directly comparable to those of Litvinenko and Somov, since
we have invoked a different field configuration. However, our particle code 
could be applied to the particular current sheet geometry invoked by these 
authors and their analytical results compared with simulations.  

\section*{Acknowledgements}
The authors are grateful to Gordon Emslie and Alec MacKinnon for helpful
discussions. BH was supported by an EPSRC CASE studentship, and the work was 
also funded partly by the UK Department of Trade and Industry.

\section*{References}

\ref{Aparicio, J., Haines, M.G., Hastie, R.J., and Wainwright, J.P.: 1998, 
Phys. Plasmas {\bf 5}, 3180.}

\ref{Aschwanden, M.J.: 2002, {\it Space Science Reviews} {\bf 101}, 1.}

\ref{Biskamp, D.: 2000, \textsl{Magnetic Reconnection in Plasmas}, Cambridge 
University Press, Cambridge, p. 36.}

\ref{Browning, P.K. and Vekstein, G.E.: 2001: {\it J. Geophys. Res.} 
{\bf 106}, 18677.}

\ref{Bruhwiler, D.L. and Zweibel, E.G.: 1992, {\it J. Geophys. Res.} {\bf 97}, 
10825.} 

\ref{Bulanov, S.V.: 1980, {\it Sov. Astron. Lett.} {\bf 6}, 206.} 

\ref{Craig, I.J.D. and Litvinenko, Yu.E.: 2002, {\it Astrophys. J.} {\bf 570},
387.}

\ref{Craig, I.J.D. and McClymont A.N.: 1991, {\it Astrophys. J.} {\bf 371}, 
L41.}

\ref{Ishihara, O. and Hirose, A.: 1981, {\it Phys. Rev. Lett.} {\bf 46}, 771.}

\ref{Kopp, R. A. and Poletto, G.: 1986, in A. Poland (ed.), \textsl{Coronal 
and Prominence Plasmas}, NASA CP 2442, p. 469.}
 
\ref{Kulsrud, R.M.: 1998, {\it Phys. Plasmas} {\bf 5}, 1599.}

\ref{Litvinenko, Y.E.: 1996, {\it Astrophys. J.} {\bf 462}, 997.}

\ref{Litvinenko, Y.E. and Somov, B.V.: 1993, {\it Solar Phys.} {\bf 146}, 127.}

\ref{Martens, P.C.H.: 1988, {\it Astrophys. J.} {\bf 330}, L131.} 

\ref{Miller, J. A.: 1998, {\it Space Science Reviews} {\bf 86}, 79.}

\ref{Mori, K., Sakai, J., and Zhao, J.: 1998, ApJ {\bf 494}, 430.}

\ref{Petkaki, P. and MacKinnon, A.L.: 1997, {\it Solar Phys.} {\bf 172}, 279.}

\ref{Priest, E.R and Forbes, T.A.: 2001, \textsl{Magnetic Reconnection: MHD
Theory and Applications}, Cambridge University Press, Cambridge, p. 180.}

\ref{Speiser, T.W.: 1965, {\it J. Geophys. Res.}, {\bf 70}, 4219.}

\ref{Strauss, H.R.: 1976, {\it Phys. Fluids}, {\bf 19}, 134.} 

\ref{Wilson, H.R., et al.: 2002, \textsl{Proceedings of the 19th IAEA Fusion 
Energy Conference}, International Atomic Energy Agency, Vienna, in press
(paper no. FT/1-5).} 

\end{document}